\documentclass[doc]{apa6}\usepackage[]{graphicx}\usepackage[]{color}
\makeatletter
\def\maxwidth{ %
  \ifdim\Gin@nat@width>\linewidth
    \linewidth
  \else
    \Gin@nat@width
  \fi
}
\makeatother

\definecolor{fgcolor}{rgb}{0.345, 0.345, 0.345}
\makeatletter
\@ifundefined{AddToHook}{}{\AddToHook{package/xcolor/after}{\definecolor{fgcolor}{rgb}{0.345, 0.345, 0.345}}}
\makeatother

\usepackage{framed}
\makeatletter
 {\par\unskip\endMakeFramed%
 \at@end@of@kframe}
\makeatother

\definecolor{shadecolor}{rgb}{.97, .97, .97}
\definecolor{messagecolor}{rgb}{0, 0, 0}
\definecolor{warningcolor}{rgb}{1, 0, 1}
\definecolor{errorcolor}{rgb}{1, 0, 0}
\makeatletter
\@ifundefined{AddToHook}{}{\AddToHook{package/xcolor/after}{
\definecolor{shadecolor}{rgb}{.97, .97, .97}
\definecolor{messagecolor}{rgb}{0, 0, 0}
\definecolor{warningcolor}{rgb}{1, 0, 1}
\definecolor{errorcolor}{rgb}{1, 0, 0}
}}
\makeatother
\newenvironment{knitrout}{}{} 

\usepackage{alltt} 

\usepackage[natbibapa]{apacite}
\usepackage[american]{babel}
\usepackage[utf8x]{inputenc}
\usepackage{csquotes}

\usepackage{amsmath}
\usepackage{amssymb}

\usepackage[doublespacing]{setspace}
\usepackage[pagewise]{lineno}

\usepackage{amsmath,amssymb,amsfonts}

\usepackage{url}   
\usepackage{graphicx}   
\usepackage{verbatim}   
\usepackage{caption}
\usepackage{subcaption}
\usepackage{pdflscape}

\usepackage{fancyvrb}
\usepackage[most]{tcolorbox}

\usepackage{newfloat}
\DeclareFloatingEnvironment[
]{listing}

\title{The posterior probability of a null hypothesis given a statistically significant result}
\shorttitle{The probability of the null being true}

\twoauthors{Daniel J. Schad}{Shravan Vasishth}

\twoaffiliations{HMU Health and Medical University, Potsdam, Germany.}{University of Potsdam, Potsdam, Germany.}

\leftheader{Schad \& Vasishth}

\authornote{Please send correspondence to danieljschad@gmail.com. ORCID, DJS: 0000-0003-2586-6823, SV: 0000-0003-2027-1994. The authors read and approved the final version of the manuscript. The authors have no conflicts of interest with respect to their authorship or the publication of this article.}

\journal{Manuscript} 
\volume{} 

\abstract{
When researchers carry out a null hypothesis significance test, it is tempting to assume that a statistically significant result lowers Prob(H0), the probability of the null hypothesis being true. Technically, such a statement is meaningless for various reasons: e.g., the null hypothesis does not have a probability associated with it. However, it is possible to relax certain assumptions to compute the posterior probability Prob(H0) under repeated sampling. We show in a step-by-step guide that the intuitively appealing belief, that Prob(H0) is low when significant results have been obtained under repeated sampling, is in general incorrect and depends greatly on: (a) the prior probability of the null being true; (b) type-I error rate, (c) type-II error rate, and (d) replication of a result. Through step-by-step simulations using open-source code in the R System of Statistical Computing, we show that uncertainty about the null hypothesis being true often remains high despite a significant result. To help the reader develop intuitions about this common misconception, we provide a Shiny app (https://danielschad.shinyapps.io/probnull/). We expect that this tutorial will help researchers better understand and judge results from null hypothesis significance tests.
}

\ccoppy{Draft of \today} 

\IfFileExists{upquote.sty}{\usepackage{upquote}}{}
\begin{document}

\maketitle

Null hypothesis significance testing (NHST), as it is practised in all areas of science, involves a fairly straightforward procedure. We begin by positing a null hypothesis $H_0$, usually a point null hypothesis that a parameter $\mu$ has a specific value: $H_0: \mu=\mu_0$. Then we collect data, compute the sample mean $\bar{x}$, and estimate the standard error $SE$ from the sample standard deviation $s$ and sample size $n$ by computing $SE=s/\sqrt{n}$. Next, we compute some statistic such as the observed t-statistic, $t_{observed}=\frac{\bar{x}-\mu_0}{SE}$. If the absolute value of the observed t-statistic is larger than the absolute value of some critical t-value, we reject the null hypothesis.  Usually we also compute the p-value, which is the probability of obtaining the observed t-statistic, or a value more extreme, assuming that the null hypothesis is true. 
Conventionally, when the p-value is less than $0.05$, we reject the null hypothesis. A common phrasing is to say that we have a ``statistically significant'' result, and that the effect of interest is ``reliable.'' As is well known, an issue that is of great importance here is false and true discovery rates \citep{betancourt2018calibrating}. 

The false discovery rate is the probability of incorrectly rejecting the null when the null is in fact true; this is referred to as type-I error rate. To make a decision on the null hypothesis, conventionally a decision threshold of $\alpha = 0.05$ is used. If the test's assumptions are satisfied within the population being examined, then as a result, the type-I error rate will match $\alpha$.

The true discovery rate is the probability of correctly rejecting the null when $\mu$ has some specific point value that is not the null value $\mu_0$; this is usually called power. The quantity (1-power) is called type-II error rate, often written as $\beta$; it is the probability of incorrectly accepting the null when it is false with some specific value for $\mu$, i.e., when the true $\mu$ is some specific point value other than $\mu_0$.  

It is well-known that power needs to be high in order to draw inferences correctly from a statistically significant result \citep{button2013power}; when power is low, statistically significant results are \textit{guaranteed} to be overestimates and can even have the wrong sign \citep{gelmancarlin}, and they are likely to be unreplicable 
\citep{VasishthMertzenJaegerGelman2018}. In other words, a significant result under low power is \textit{never} ``reliable'' in any sense of the word.

Researchers sometimes assume that a significant result reduces \textit{the probability of the null hypothesis being true} to a low value. 
For example, a survey by \cite{tam2018doctors} reports this widespread misunderstanding of the p-value by medical doctors. As they put it:

\begin{quote}
Many respondents conceptualised the P value as numerically indicating the natural probability of some phenomenon --- for instance, a 95\% or 5\% chance of the truth or falsity of a hypothesis in the real world.
\end{quote}

A second example comes from \cite{doherty2002fluoride}. On page 376, Table 2, they state that ``[the p-value] is the probability that the null hypothesis is true.'' A third example comes from a textbook written for medical researchers 
\citep{harristaylor}. On their page 24, they write: 
``The P value is used when we wish to see how likely it is that a hypothesis is true''; and on 
page 26, they write: ``The P value gives the probability that the null hypothesis is true.''

Here, we investigate the posterior probability of the null hypothesis being true under different possible assumptions. Specifically, we consider cases where power is low, medium, or high, and when type-I error rate is 0.05, 0.01, or 0.005. Moreover, we look at how replication studies affect the posterior probability of the null hypothesis.

Intuitively, it does seem obvious that rejecting the null hypothesis after finding a significant result leads us to believe that the posterior probability of the null hypothesis being true is low. We will show in this paper that this intuitive belief is in general wrong \citep[see also][]{button2013power, bayarri2016rejection, colquhoun2014investigation, xxx2005most}, but that instead, given a significant result, the posterior probability for the null hypothesis widely varies between rather small values (when statistical power is high, greater than 0.80, and type-I error rate is low, 0.005 - two extreme situations that are rarely or never realized) and larger values (when power is low or type-I error rate is higher, 0.05).

Note that, technically, talking about ``the probability that the null hypothesis is true'' is meaningless in the NHST framework. Probability mass functions can only be associated with discrete outcomes that constitute a random variable. An example from everyday life would be the probability of catching a train when running late: there are two possible outcomes, either one gets the train or not, and each outcome has a probability associated with it. By contrast, the null hypothesis is not a random variable and therefore cannot have a probability associated with the two possible outcomes of being true or false. The null hypothesis is either true or it is false. 

However, in order to talk about the probability of the null hypothesis being true or false, we can assume for the moment that the null hypothesis \textit{is} a random variable.  Suppose that before running the experiment, we begin with the assumption that the null hypothesis is believed to be true with some probability $\theta_{prior}$. Once we get a significant result, the probability of the null being true should (intuitively) fall to some low value $\theta_{posterior}$. 

\begin{figure}[!h]
\centering
\begin{knitrout}
\definecolor{shadecolor}{rgb}{0.969, 0.969, 0.969}\color{fgcolor}
\includegraphics[width=\maxwidth]{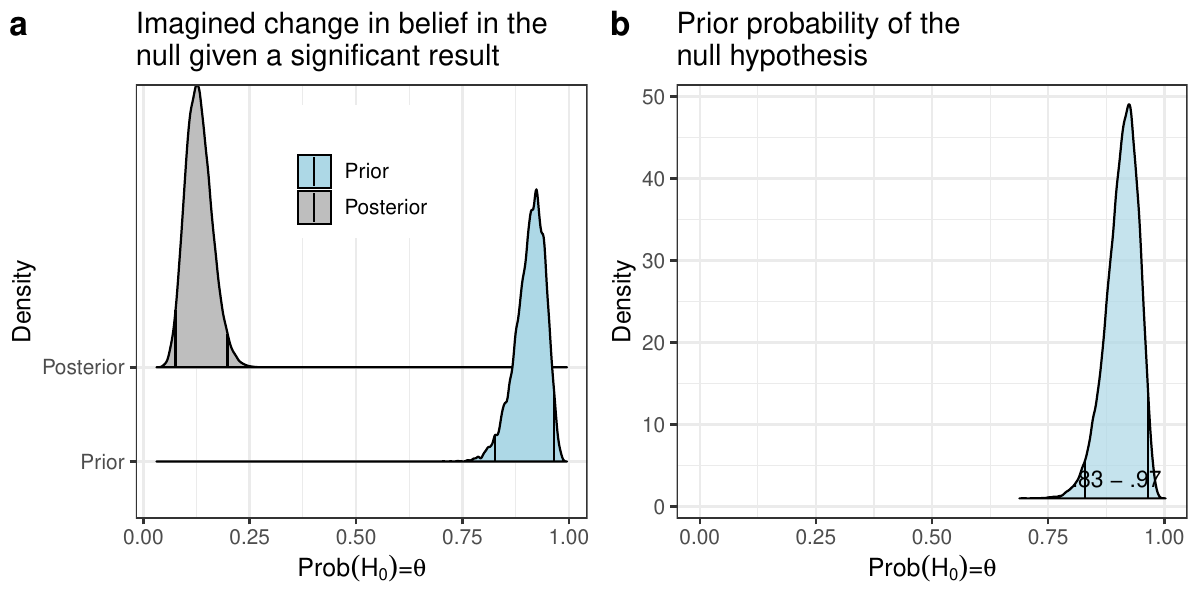} 
\end{knitrout}
\caption{Prior and posterior probability of the null hypothesis. a) An illustration of how our belief in the null hypothesis---expressed as a probability distribution--- might hypothetically shift once we see a statistically significant result. The vertical lines show the 95\% credible intervals. 
b) Prior probability of the null hypothesis being true, expressed as a Beta(60,6) distribution.}
\label{fig:illustration}
\end{figure}

But these point values $\theta_{prior}$ and $\theta_{posterior}$ only partly characterize our beliefs. Before running the experiment, we surely have some uncertainty about the probability $\theta_{prior}$ that the null is true. For example, the null may be true at the outset with probability somewhere between 85 and 95\%. Thus, the prior probability of the null being true has a probability \textit{distribution} associated with it, it cannot be just a \textit{point} value. Explicitly capturing this uncertainty in a distribution, which we do here, goes beyond previous treatments of this topic \citep{button2013power, bayarri2016rejection, colquhoun2014investigation, xxx2005most}. Still relying on intuition, we might say that a statistically significant result should shift this probability distribution to some low range, say 10-20\%. Such a hypothetical situation is visualized in Figure \ref{fig:illustration}a.

To summarize, strictly speaking, the null hypothesis is either true or false, it has no probability distribution associated with it. So, one cannot even talk about the probability of the null hypothesis being true. Nevertheless, it is possible to relax this strict stipulation and ask ourselves: how strongly do we believe that the null is true before and after we do a significance test?
A domain expert working in a particular field should be able to state, as a probability distribution, his or her a priori confidence level in a particular null hypothesis. In practice, elicitation from an expert might be required \citep{ohagan2006uncertain,OakleyOHagan}. 

Bayes' rule allows us to calculate this posterior probability of the null hypothesis being true.  Bayes' rule states that, given a vector of data $y$, we can derive the probability density function of a parameter or a vector of parameters $\theta$ given data, $f(\theta\mid y)$, 
by multiplying the likelihood function of the data, $f(y \mid \theta)$, with the prior probability of the parameter(s), $f(\theta)$, and dividing by the marginal likelihood of the data, $f(y)$:

\begin{equation}
f(\theta\mid y) = \frac{f(y \mid \theta) f(\theta)}{f(y)}
\end{equation}

The marginal likelihood of the data can be computed by integrating out the parameter(s) $\theta$:

\begin{equation}
f(y) = \int f(y\mid \theta) f(\theta)\, d\theta
\end{equation}

As \cite{mcelreath2016statistical} mentions, 
Bayes' rule can be used to work out the posterior probability of the null being true given a significant effect under repeated sampling \citep{button2013power, bayarri2016rejection, colquhoun2014investigation, xxx2005most}. 
This is what we turn to next. Before we can carry out this computation, 
we have to decide on the prior probability of the null hypothesis being true; this is not too difficult to determine for specific research questions. We could start by  eliciting from a researcher their prior belief about the probability of some particular null hypothesis being true: $Prob(H_0~true)$. 
Given such a prior probability for the null hypothesis, we then stipulate a type-I error rate, $Prob(sig | H_0~true)=\alpha$ and a type-II error rate, $Prob(not~sig | H_0~false)=\beta$. (When we write $H_0~false$, we mean that the null is false with some specific value for the parameter $\mu$).
Once we have these numbers, 
Bayes' rule allows us to compute $Prob(H_0~true|sig)$, the posterior probability of the null being true given a significant result under repeated sampling. In other words, Bayes' rule helps us quantify the extent to which our prior belief should shift in the light of the long-run probability of obtaining a statistically significant result:

\begin{equation}
Prob(H_0~true | sig) = \frac{Prob(sig|H_0~true)\times Prob(H_0~true)}{Prob(sig)}
\end{equation}

The denominator $Prob(sig)$ is the marginal probability of obtaining a significant effect under repeated sampling, and is straightforward to compute using the law of total probability \citep{kolmogorov2018foundations}. This law states that the probability of a random variable Z, $Prob(Z)$, given another random variable A, is $Prob(Z | A)Prob(A) + Prob(Z | \neg A) Prob(\neg A)$. Translating this to our particular question, the event $Z$ is the significant effect we obtained under repeated sampling, and the event $A$ is the null hypothesis being true or false.

\begin{equation}
\begin{split}
Prob(sig) =& Prob(sig |  H_0~true) Prob(H_0~true) + Prob(sig | H_0 ~false) Prob(H_0~false) \\
        =& \alpha \times Prob(H_0~true) + (1-\beta) \times (1-Prob(H_0~true)) \\
        =& \alpha \times \theta + (1-\beta) \times (1-\theta) \\
\end{split}
\end{equation}

\noindent
The last line above arises because 
$Prob(sig |  H_0~true)= \alpha$ (type-I error rate), $Prob(sig | H_0 false) = 1-\beta$ (power), and $Prob(H_0~true)=\theta$ (the prior probability of the null being true).

Taken together, we can compute the probability for the null hypothesis given a significant effect as:

\begin{equation}
Prob(H_0~true | sig) = \alpha\times \frac{\theta}{(\alpha\times \theta + (1-\beta)\times(1-\theta))}
\end{equation}

Thus, $Prob(H_0~true | sig)$ is really a function of three quantities:

\begin{enumerate}
\item The false discovery rate, or type-I error rate $\alpha$.
\item The true discovery rate, or power $(1-\beta)$, where $\beta$ is type-II error rate.
\item The prior probability of null being true ($\theta$).
\end{enumerate}

We will now compute, under different assumptions, the posterior probability of the null being true given a significant effect. 
Before we can do this, we have to decide on a prior probability of the null hypothesis being true. What is a reasonable prior distribution to start with?
In a recent paper, \cite{benjamin2018redefine} write the following:
``Prediction markets and analyses of replication results both suggest that for psychology experiments, the prior odds of H1 relative to H0 may be only about 1:10. A similar number has been suggested in cancer clinical trials, and the number
is likely to be much lower in preclinical biomedical research.'' 
A prior odds of 1:10 of the alternative being true relative to the null means that the probability of the null being true is about 90\%. We take this estimate as a starting point; below we will also consider alternative scenarios where the probability of the null being true is lower. 

For now, we will assume that the null hypothesis $H_0$ has the high prior probability of 90\% of being true. Just as a coin has heads and tails as possible outcomes, the null hypothesis can have two possible outcomes, true or false, each with some probability.   Thus, we can now talk about the probability  $\theta$ of the null hypothesis being true. We can model this by assuming that a success or failure is generated from a Bernoulli process that has probability of success $\theta$:

\begin{equation}
H_0 \sim Bernoulli(\theta)
\end{equation}

Because our prior belief that the null is true will come with some uncertainty (it is not merely a point value), we can model this prior belief through a Beta distribution. 
We chose a Beta distribution here because the prior on a probability parameter is usually a Beta distribution (for example, consider the classic Beta-Binomial conjugate case in Bayesian statistics).
For example, a prior $Beta(60,6)$ on $\theta$ expresses the assumption that the prior probability of the null being true is between 0.83 and 
0.97 with probability 95\% (approximately), with mean 
probability approximately 0.90. 
The lower and upper bounds of the 95\% credible interval can be computed using the inverse cumulative distribution function of the Beta distribution. We simply solve the integrals for the lower and upper bounds:

\begin{equation}
\int_{-\infty}^{lower} f(x) dx = \int_{-\infty}^{lower} \frac{\Gamma(a+b)}{\Gamma(a)\Gamma(b)}x^{a-1}(1-x)^{b-1}\, dx = 0.025
\end{equation}

and 

\begin{equation}
\int_{-\infty}^{upper} f(x) dx = \int_{-\infty}^{upper} \frac{\Gamma(a+b)}{\Gamma(a)\Gamma(b)}x^{a-1}(1-x)^{b-1}\, dx = 0.975
\end{equation}

The mean can be computed  from the fact that a random variable $X$ that is generated from a Beta distribution with parameters 
$a$ and $b$ has expectation:

\begin{equation}
E[X] = \frac{a}{a + b}
\end{equation}

Given the $a$ and $b$ parameters, the expectation is:

\begin{equation}
E[X] = \frac{a}{a + b} = \frac{60}{60+6} = 0.9
\end{equation}

Figure \ref{fig:illustration}b 
visualizes this prior probability of the null hypothesis being true.


Given this prior probability density function Beta(60,6) for $\theta$, we are now in a position to investigate how the posterior probability of the null being true changes under different assumptions. 
We can use Monte Carlo sampling to compute the posterior probability of $H_0$ being true given significant results under repeated sampling:

\begin{enumerate}
\item
Fix $\alpha$ (type-I error rate) and $\beta$ (type-II error rate).
\item Do 100,000 times:
\begin{enumerate}
\item Sample one value $\theta$ from the Beta(60,6) distribution.
\item Compute the posterior probability of $\theta$ given  $\alpha, \beta$, and the sampled value of $\theta$ from the prior distribution Beta(60,6):

\begin{equation}
\alpha\times \frac{\theta}{(\alpha\times \theta + (1-\beta)\times(1-\theta))}
\end{equation}
\item 
Store this posterior probability of the null hypothesis being true.

\end{enumerate}
\item Plot the distribution of the stored probabilities, or display summary statistics such as the mean and the 95\% credible interval.
\end{enumerate}

There are two interesting cases. The first is when statistical power is low (30\%); we will show that in this case, it simply doesn't matter much whether we get a significant result or not.
The posterior probability of the null being true will not change substantially; this is regardless of whether type-I error rate is 0.05 or some lower value like 0.005, as recommended by \cite{benjamin2018redefine}. Note that here, when we say that the posterior probability of the null doesn't change substantially, we are not talking about the mean of the posterior probability; rather, we are more interested in the 95\% credible interval, which represents the uncertainty of the posterior probability. Thus, even if the mean of the posterior distribution is numerically lower than the mean of the prior distribution, if the uncertainty associated with the posterior distribution is high, the prior-to-posterior change in mean probability is not really meaningful.

The other case is where statistical power is high (90\%); here, the posterior probability of the null being true will change considerably once we have a significant result under repeated sampling, especially if we follow the recommendation of \cite{benjamin2018redefine} to lower type-I error rate to 0.005.  

Anticipating our main conclusion, when the prior probability of the null being true is low, 
the only situation where a statistically significant effect under repeated sampling can shift our belief substantially against the null hypothesis being true is when statistical power is high. When power is low, it simply doesn't matter whether you lower the type-I error rate to 0.005, as suggested by \cite{benjamin2018redefine} and others.

Null hypothesis significance testing only makes sense if power is high; in all other situations, the researcher is wasting their time computing p-values. When power is low, the intuition, that a significant result will lead to a low posterior probability of the null hypothesis being true, is an illusion.

\begin{figure}[!htbp]
\centering
\begin{knitrout}
\definecolor{shadecolor}{rgb}{0.969, 0.969, 0.969}\color{fgcolor}
\includegraphics[width=\maxwidth]{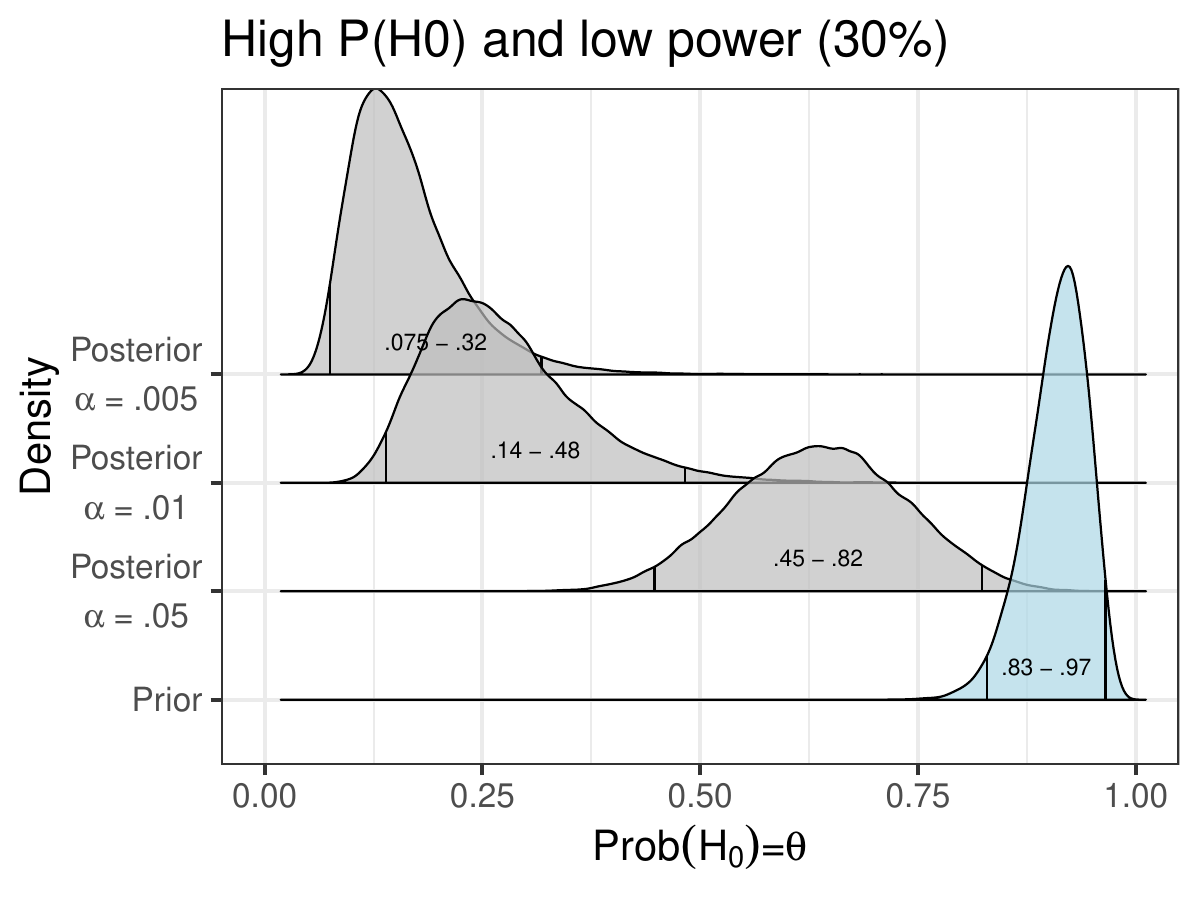} 
\end{knitrout}
  \caption{Probabilities for the Null-hypothesis, $P(H_0)=\theta$. Prior probability (blue) and posterior probabilities given a significant effect (grey) at a type-I error rate threshold $\alpha$ of $0.05$ and $0.01$. This is shown for a situation of low statistical Power of $30$\% (type-II error rate $\beta = 70$\%) and with a high prior probability for the null hypothesis ($\theta \sim Beta(60,6)$; blue).}\label{fig:postProb1}
\end{figure}

\section*{Case 1: Investigating the posterior probability of the null hypothesis being true when power is low (type-II error rate 0.70) and the prior probability of the null hypothesis is high (Mean Prob(H0)=.90)}

We first look at the posterior probability for different situations: As a first case, we investigate the posterior probability of the null hypothesis being true when the prior probability of the null hypothesis is high (Mean Prob(H0)=.90) and when power is low (type-II error rate 0.70). We investigate several scenarios by using different type-I error rates ($\alpha = 0.05, 0.01$ and $0.005$).

\paragraph{Scenario 1: Low power (0.30), type-I error rate 0.05}
Let type-I error rate be $\alpha = 0.05$ and type-II error rate be $\beta = 0.70$. So, we have power at $1-\beta=0.30$. Such low power is by no means an uncommon situation in areas like cognitive psychology, psycholinguistics, and linguistics; examples from psycholinguistics are discussed in \cite{JaegerEngelmannVasishth2017,jager2020interference,VasishthMertzenJaegerGelman2018,SampleSizeCBB2022}. In these examples, power is quite low, even for repeated measures designs, where stimuli are repeatedly presented to an individual. 

Figure \ref{fig:postProb1} shows the prior and posterior distributions. 
The prior distribution is plotted in blue.
Figure \ref{fig:postProb1} shows that getting a significant result hardly shifts our belief regarding the null. This should be very surprising to researchers who believe that a significant result shifts their belief about the null hypothesis being true to a low value.
Next, consider what happens when we reduce type-I error rate to 0.01, which is lower than the traditional 0.05. 

\paragraph{Scenario 2: Low power (0.30), type-I error rate 0.01}
Many researchers \citep{benjamin2018redefine} have suggested that lowering type-I error rate will resolve many of the problems with NHST. Let's start by investigating what changes when we decrease type-I error rate to $0.01$ (researchers like \citealp{benjamin2018redefine} have proposed 0.005 as a threshold for type-I error rate; we turn to this proposal below). Type-II error rate is held constant at 0.70.

Figure \ref{fig:postProb1} shows that 
lowering type-I error rate does shift our posterior probability of the null being true a bit more but not enough to have any substantial effect on our beliefs. It seems unreasonable to discard a null hypothesis if the posterior probability of it being true lies between 14 and 48\%. 

This result highlights a problem that has been discussed before, namely that p-values may be a valuable way to quantify information, but should not be used to perform decisions on the null versus the alternative hypotheses \citep{amrhein2019inferential}. Such decisions can be highly problematic if the posterior probability for the null hypothesis is still quite high after a significant result.

\section*{Case 2: Incorporating uncertainty about type-II error rate}

\begin{figure}[!htbp]
\centering
\begin{knitrout}
\definecolor{shadecolor}{rgb}{0.969, 0.969, 0.969}\color{fgcolor}
\includegraphics[width=\maxwidth]{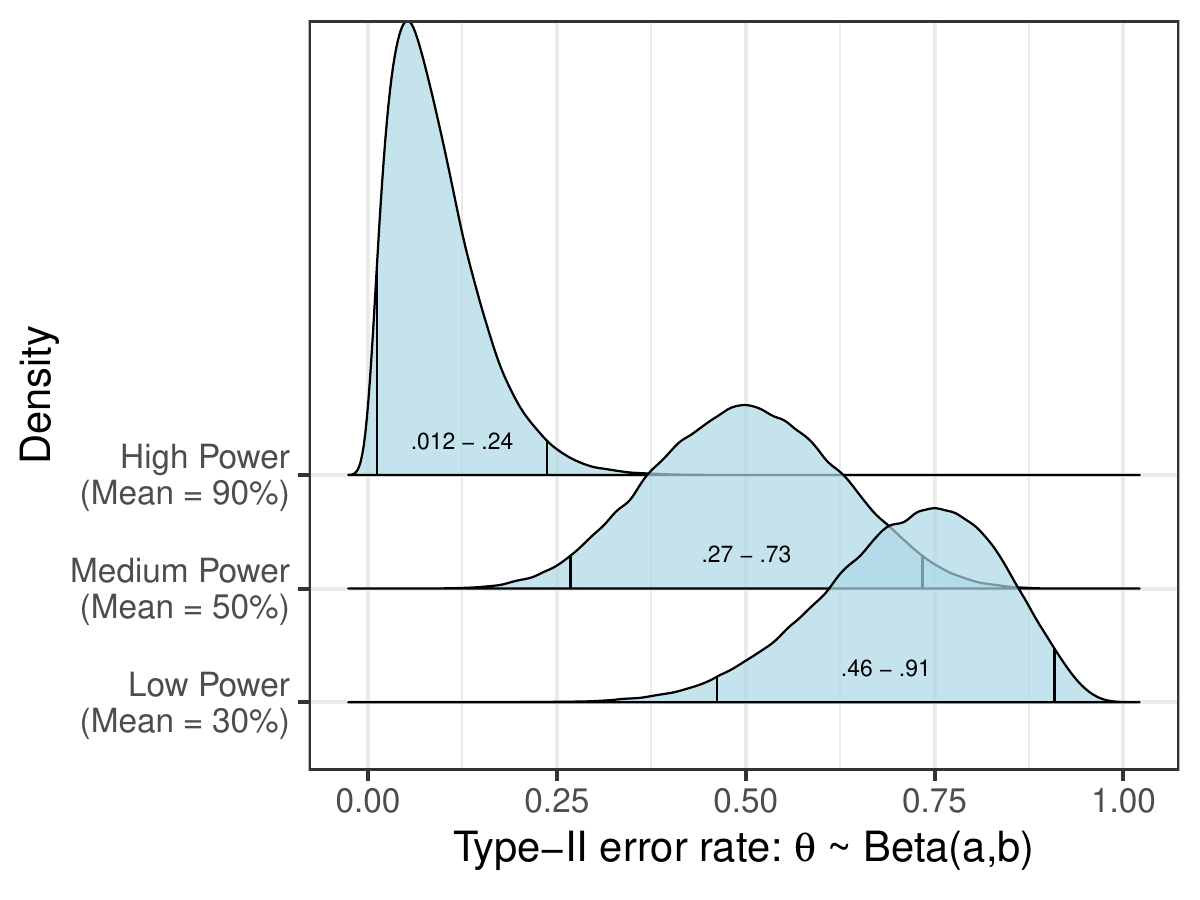} 
\end{knitrout}
\caption{Visualization of the probability distribution associated with type-II error rate $\beta$ corresponding to low power ($\beta \sim Beta(10,4)$, mean power $E[1-\theta] = 30$\%), medium power ($\beta \sim Beta(8,8)$, mean power $E[1-\theta] = 50$\%), and high power ($\beta \sim Beta(2,20)$, mean power $E[1-\theta] = 90$\%) . Recall that Power is 1-type-II error rate.}\label{fig:type2}
\end{figure}

So far, we have been assuming a point value as representing power. 
However, power is really a function that depends (inter alia) on the magnitude of the true (unknown) effect. Power therefore also has some uncertainty associated with it, because we do not know the magnitude of the true effect, and we do not know the true standard deviation.
We can introduce uncertainty about power (or equivalently, uncertainty about type-II error rate) into the picture by setting our prior on $\beta \sim Beta(10,4)$, so that the type-II error rate is around 70\%. Different levels of power (1-type-II error rate) are visualized in Figure \ref{fig:type2}, and the low power situation of 30\% is shown in the bottom row of the figure.

\paragraph{Scenario 3: Low power, type-I error rate 0.05}

\begin{figure}[!htbp] 
\begin{knitrout}
\definecolor{shadecolor}{rgb}{0.969, 0.969, 0.969}\color{fgcolor}
\includegraphics[width=1.0\textwidth]{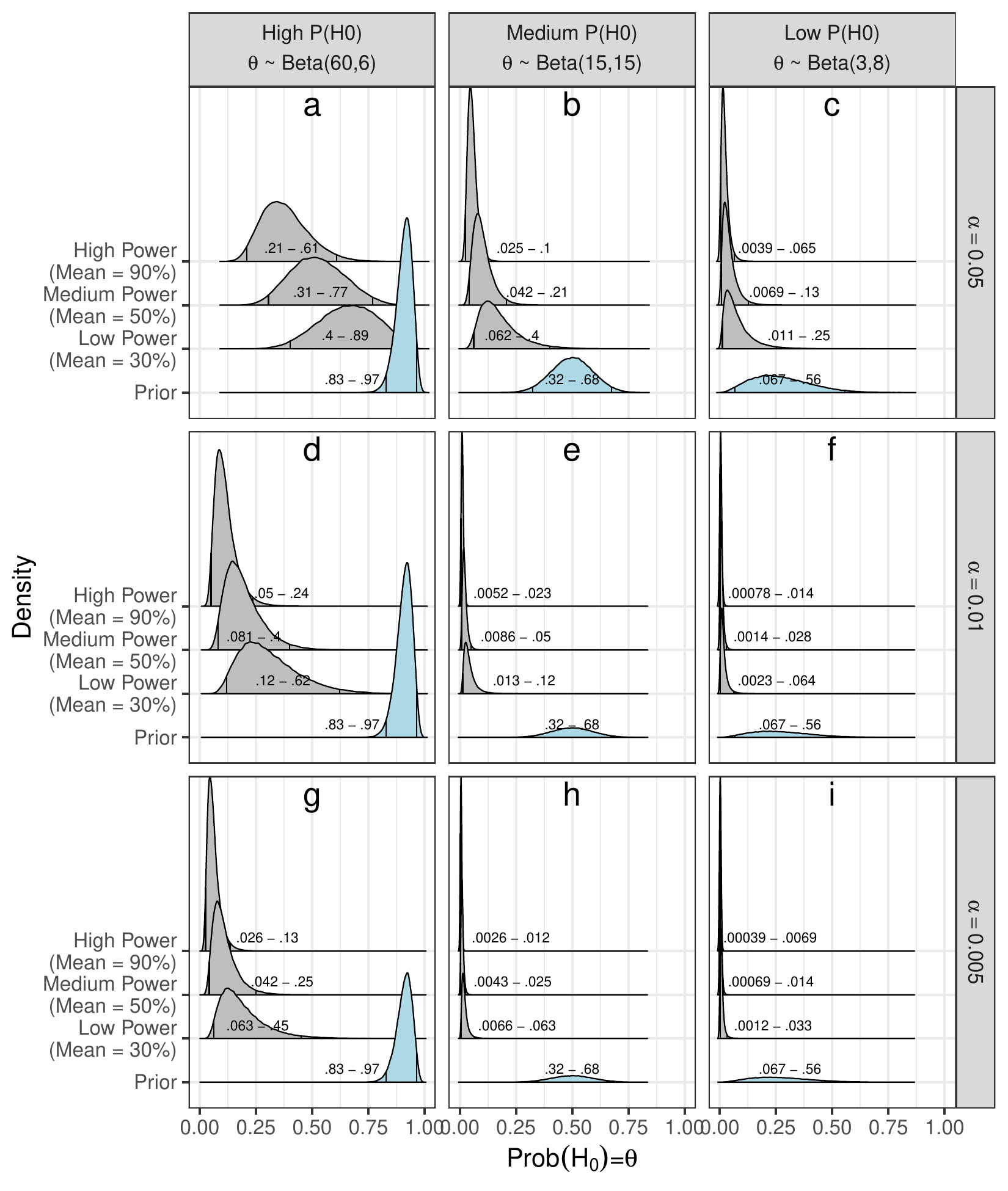} 
\end{knitrout}
\caption{Probabilities for the null hypothesis, $P(H_0)=\theta$, considering uncertainty about power. Prior probability (blue) and posterior probabilities given a significant effect (grey) at a type-I error rate $\alpha$ of $0.05$ (upper panels), $0.01$ (middle panels), and $0.005$ (lower panels). This is shown for situations of low statistical Power, $\beta \sim Beta(10,4)$ (mean type-II error rate of about $\beta = 70$\%, mean Power of about $30$\%), medium statistical Power, $\beta \sim Beta(8,8)$ (mean type-II error rate of $\beta = 50$\%, mean Power of $50$\%), and high statistical Power, $\beta \sim Beta(2,20)$ (mean type-II error rate of about $\beta = 10$\%, mean Power of about $90$\%), and for situations with a high (left panels), medium (middle panels), and low (right panels) prior probability for the null hypothesis.}\label{fig:postProb}
\end{figure}

Incorporating the uncertainty about type-II error rate (equivalently, power) increases the uncertainty about the posterior probability of the null quite a bit. Compare Figure \ref{fig:postProb1} ($\alpha = 0.05$) and 
Figure \ref{fig:postProb}a (low power). 
Figure \ref{fig:postProb}a
shows that the posterior of the null being true now lies between 40 and 90\% (as opposed to 45 and 82\% in Figure \ref{fig:postProb1}). 

\paragraph{Scenario 4: Type-I error rate 0.01}
Having incorporated uncertainty into type-II error rate, consider now what happens if we lower type-I error rate to 0.01, from 0.05. Figure \ref{fig:postProb}d shows (cf. low power) that now the posterior distribution for the null hypothesis shifts to the left quite a bit more, but with wide uncertainty (10-60\%). Even with a low type-I error rate of 0.01, we should be quite unhappy rejecting the null if the posterior probability of the null being true is between the wide range of 10 and 60\%. 

\paragraph{Scenario 5: Type-I error rate 0.005}
Next, consider what happens if we lower type-I error rate to 0.005. This is the suggestion from \cite{benjamin2018redefine}. Perhaps surprisingly, Figure \ref{fig:postProb}g shows that now the posterior distribution for the null hypothesis does not shift much compared to Scenario 4 (see Fig. \ref{fig:postProb}d): the range is 6 to 45\% (compare with the range 10-60\% in scenario 4). 
Thus, when power is low, there is simply no point in engaging in null hypothesis significance testing. Simply lowering the threshold of type-I error rate to 0.005 will not change much regarding our belief in the null hypothesis.

\section*{Case 3: Investigating the posterior probability of the null hypothesis being true when power is high}

As a second case, we investigate the posterior probability of the null hypothesis being true when power is high.
We consider the case where power is around 90\%. We will still assume a high prior probability for the null (Mean Prob(H0) = .90). We will consider three scenarios: type-I error rate at 0.05, 0.01, and 0.005.

\paragraph{Scenario 6: High power (0.90), type-I error rate 0.05}
First consider a situation where we have high power and type-I error rate is at the conventional 0.05 value. The question here is: in high power situations, does a significant effect shift our belief in any substantial manner away from the null, with type-I error rate at the conventional value of 0.05?
The prior on type-II error rate is shown in Figure \ref{fig:type2}. The mean type-II error rate is 10\%, implying a mean for the power distribution to be 90\%.
Perhaps surprisingly, Figure \ref{fig:postProb}a shows that even under high power, our posterior probability of the null being true does not shift dramatically: the probability lies between 20 and 60\%.

\paragraph{Simulation 7: High power (mean 0.90), type-I error rate 0.01}

Next, we reduce type-I error rate to 0.01.
Figure \ref{fig:postProb}d shows that when power is high and type-I error rate is set at 0.01, we get a big shift in posterior probability of the null being true: the range is 5-25\%.

\paragraph{Simulation 8: High power (mean 0.90), type-I error rate 0.005}

Next, in this high-power situation, we reduce type-I error rate to 0.005.
Figure \ref{fig:postProb}g shows that when power is high and type-I error rate is set at 0.005, we get a decisive shift in posterior probability of the null being true: the range is now 2-13\%.

\section*{Cases 3 and 4: Prior probability for the null is medium or low}

Next, we consider cases where the prior probability for the null is medium or low.

\paragraph{Low prior probability for the null: Mean Prob(H0)=.10}

One possible objection to the above analyses is that the prior probability for the null hypothesis could often be much smaller than an average of 90\%. Indeed, in some situations, the null hypothesis may be very unlikely. We here simulate a situation where the prior probability for the null is an average of 10\% ($\theta \sim Beta(3,8)$). For this situation, Figures \ref{fig:postProb}c,f, and i show that the posterior probability for the null is always decisively low. Even for a conventional type-I error rate of $\alpha = 0.05$ in a low-powered study, the posterior probability for the null ranges from 1 to 25\%, which is quite low, and when turning to smaller $\alpha$ levels or higher power, the effect is decisive.

However, this is of course not very informative, as we started out assuming that the null hypothesis was unlikely to be correct in the first place. Thus, we haven't learned much; a statistical significance test would just confirm what we already believed with high certainty before we carried out the test.

\paragraph{Medium prior probability for the null: Mean Prob(H0)=.50}

Now consider the case where the prior probability for the null being true lies at an average of 50\% (e.g., $\theta \sim Beta(15,15)$). Here, we don't know whether the null or the alternative hypothesis is true a priori, and both outcomes seem equally likely. In this situation, when we use a conventional type-I error rate level of $\alpha = 0.05$ in a low-powered study, a significant effect will bring our posterior probability for the null only to a range of 6-40\%, and will thus leave us with much uncertainty after obtaining a significant effect. 

However, either using a stricter type-I error rate level (e.g., $\alpha = 0.005$) or running a high-powered study each suffices to yield informative results: For a high-powered study and $\alpha = 0.05$, a significant result will (under our assumptions) bring the posterior probability to 2-10\% (Figure \ref{fig:postProb}b), which is quite informative. And for a type-I error rate level of $\alpha = 0.005$ a significant effect brings decisive evidence against the null for all the levels of power that we investigated (Figure \ref{fig:postProb}h), with a posterior probabilty of 0.7-6\% even for low-powered studies. This suggests that when the prior probabilities of the null and the alternative hypotheses are each at 50\%, then either high power or a strict type-I error rate of $\alpha = 0.005$ will yield informative outcomes once a significant effect is observed.


\section*{How the posterior probability changes with replication studies}

One topic that is often brought up when discussing the trustworthiness of research findings is replication. Indeed, it is widely accepted that replicating a research finding or effect in an independent sample provides a lot of support for it. In the present framework, we can quantify how much support this provides. Specifically, in Bayesian analyses it is possible to use the posterior from one analysis or data set as the prior for the next analysis/data set, i.e., for a replication study. The posterior of the second study then incorporates all the knowledge obtained from both results, and thus this yields an intuitive way to accumulate knowledge across studies.

\begin{figure}[!htbp] 
\begin{knitrout}
\definecolor{shadecolor}{rgb}{0.969, 0.969, 0.969}\color{fgcolor}
\includegraphics[width=1.0\textwidth]{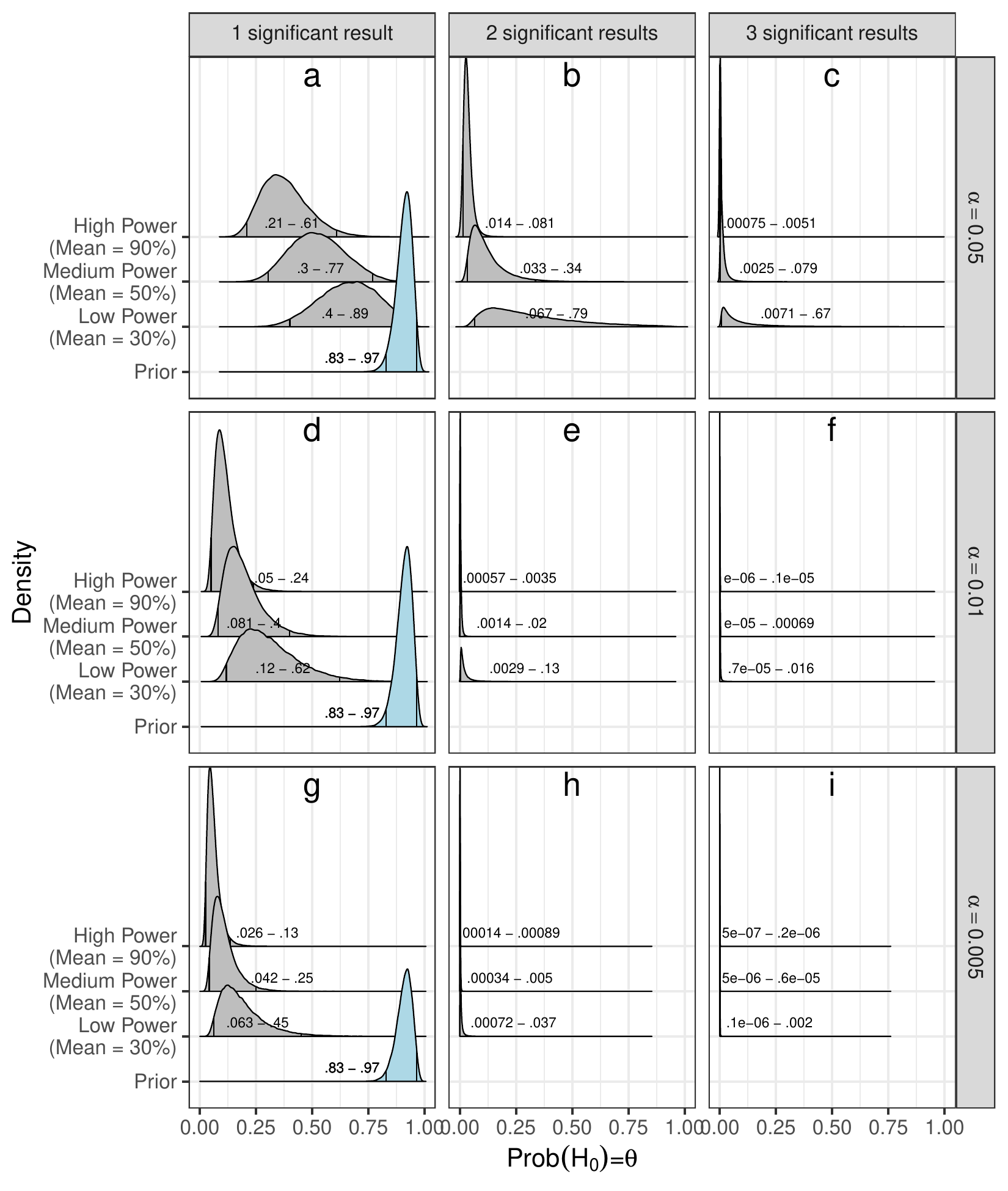} 
\end{knitrout}
\caption{Probabilities for the null hypothesis, $P(H_0)=\theta$, for different number of significant replication studies. Left panels: One study yields a significant result. Middle panels: After an original study turned out significant, one replication study with identical experimental design and significance level was also significant. Right panels: One original and two replication studies with identical designs and significance levels yielded significant results. Shown are the prior probability (blue) and posterior probabilities given a significant effect (grey) at a type-I error rate $\alpha$ of $0.05$ (upper panels), $0.01$ (middle panels), and $0.005$ (lower panels). The priors for replication studies are defined as the posterior from the previous study. Quantities are shown for situations of low statistical Power, $\beta \sim Beta(10,4)$ (mean type-II error rate of about $\beta = 70$\%, mean Power of about $30$\%), medium statistical Power, $\beta \sim Beta(8,8)$ (mean type-II error rate of $\beta = 50$\%, mean Power of $50$\%), and high statistical Power, $\beta \sim Beta(2,20)$ (mean type-II error rate of about $\beta = 10$\%, mean Power of about $90$\%).}\label{fig:repProb}
\end{figure}

Here, we followed this procedure by investigating how the posterior probability for the null hypothesis changes when a study is replicated with the identical experimental design (i.e., same power, $1 - \beta$) and with the same significance level ($\alpha$ error). Figure \ref{fig:repProb} displays the posterior probability for the null hypothesis after an original study showed a significant result (Figure \ref{fig:repProb}a, d, g; these results are identical to Figure \ref{fig:postProb}). These posteriors are then each used as a prior for a replication study. The posterior distributions from the replication studies are shown in Figure \ref{fig:repProb}b, e, h. They show that if the original and a replication study both showed a significant effect, then the posterior probability for the null hypothesis is quite low for most designs and significance levels. More precisely, the posterior probability for the null is low when either statistical power is high or when the type-I error rate is small ($\alpha = .01$ or $\alpha = .005$). However, interestingly, when power is low and $\alpha = .05$, even replicating an original effect yields a posterior probability of the null hypothesis ranging between $0.067$ and $0.79$. Thus, although the effect was replicated the posterior probability for the null is still far from decisive. This shows that low power studies really do not provide a lot of evidence for a null hypothesis, even when performing a replication study.

The picture looks more or less similar when a second replication study is conducted and turns out significant (see Figure \ref{fig:repProb}c, f, i). The main difference is that the posterior probability for the null hypothesis is now even smaller for all studied scenarios. Interestingly, for a situation of low power and an $\alpha$-level of $.05$, even a second significant replication result does not bring down the posterior probability for the null substantially. Instead, in this situation the posterior probability still ranges between $0.0071$ and $0.67$. Thus, with low power studies, even three significant results in a row are not very informative concerning the null hypothesis. When using medium/high power studies or stricter significance thresholds ($\alpha = .01$ or $\alpha = .005$), three significant results are decisive.

As an important qualification, replications are only going to yield unbiased results if there is no publication bias, i.e., if failed replications are also made public \citep{VasishthMertzenJaegerGelman2018}. Within psycholinguistics, we are aware of at least one case where a failed replication was never made public. Due to publication bias, estimates from replications could end up being biased, and could lead to misleading posterior probabilities of the null hypothesis.

\section*{Discussion}

In summary, we analyzed the posterior probability for the null given a significant effect. We provide a shiny app (https://danielschad.shinyapps.io/probnull/) that allows the user to compute the posterior distribution for different choices of prior, type-I and type-II error rates.
For psychology and preclinical biomedical research, the prior odds of H1 relative to H0 are estimated to be about 1:10 \citep{benjamin2018redefine}, reflecting a high prior probability of the null of 90\%. 
Indeed, in science, making obvious predictions may not be a valuable endeavour \citep{lykken1968statistical}, since obvious predictions may be derived from or be in line with many theoretical positions or may result from many different influential factors. Instead, testing provocative predictions, that is, predictions that no one would bet on (e.g., with a prior probability for the null hypothesis of 90\%), is what generates progress in scientific understanding, if the data forces scientists to change their thinking.

For this common and standard situation in psychology and other areas, where the null hypothesis is unlikely a priori, when power is low, the posterior probability of the null being true doesn't change in any meaningful way after seeing a significant result, even if we change type-I error rate to 0.005. This result is in line with prior suggestions that p-values should not be used to make decisions on the null or alternative hypothesis \citep{amrhein2019inferential}.
What shifts our belief in a meaningful way is reducing type-I error rate to say 0.005 (as suggested by \citealp{benjamin2018redefine} and others), \textit{as well as} running a high powered study, followed by a high-powered replication. Only this combination of high power and small type-I error rate yields informative results.

One might object here that we set the prior probability of the null hypothesis being true at an unreasonably high value. This objection has some merit; although typically the prior probability for the null may lie at 10\%, there may well be some situations where the null is unlikely to be true a priori. 
In this situation, our results show that a significant effect does indicate a very low posterior probability of the null. This is the case across a range of type-I error rate levels ($\alpha$ of 0.05, 0.01, 0.005) and for different levels of power (Figure \ref{fig:postProb}c+f+i). Even for low power studies with $\alpha = 0.05$ the posterior probability is between 1 and 25\%, which is quite low. 
So yes, if the prior probability of the null being true is already low, even with relatively low power and the standard type-I error rate level of 0.05, we are entitled to changing our belief quite strongly against the null once we have a significant effect. An obvious issue here is that if we already don't believe in the null before we do the statistical test, why bother to try to reject the null hypothesis? 
Even if we were satisfied with rejecting a null hypothesis we don't believe in in the first place, running low power studies is always a bad idea because of type M and S error.\footnote{Significant results have a higher chance of publication, i.e., published research goes through a significance filter. In this situation, low-power studies will only turn out significant when the effect is very large by chance, leading to type M(agnitude) error, i.e., an overestimation of effect sizes in the published literature. Moreover, some published effects will have the wrong sign, indicative of type S(ign) error.} As \cite{gelmancarlin} and many others before them have pointed out, significant effects from low power studies will have exaggerated estimates of effects and could have the wrong sign of the effect. The probability of the null hypothesis being true is not the only important issue in a statistical test; accurate estimation of the parameter of interest and quantifying our uncertainty about that estimate are equally important.

Replication of results is generally viewed as strong evidence against the null hypothesis and for the robustness of an effect. In Bayesian analysis, results from several replication studies can easily be combined by using the posterior from one study as the prior for the next. Using this approach, we could quantify how a succession of significant results in identical replication studies can bring down the posterior probability for the null, when the prior for the null is high. Our results showed that when using either high-power studies or low type-I error rate levels ($\alpha = .01$ or $\alpha = .005$), a significant replication makes the null hypothesis very unlikely. However, interestingly, when using a low-power design (with average power of $30$\%) and $\alpha = .05$, then even three studies with a significant effect (i.e., one original studies and two replications) still do not provide decisive evidence against the null hypothesis (the posterior probability of the null ranges between $0.0071$ and $0.67$). Thus, our results support the general view that replications provide key information about the null hypothesis, but also show that this does not hold true for low power studies with conventional type-I $\alpha$ error levels.

Last, we note that results from empirical studies usually have more than one possible explanation \citep{lykken1968statistical}. Therefore, new, surprising and unlikely predictions should be generated by a theory and tested experimentally, and alternative explanations should be ruled out \citep["strong inference"][]{platt1964strong}. In any case, a discovery, especially a surprising discovery, should not be claimed based on a single p-value.

In summary, we investigated the intuitive belief held by some researchers that finding a significant effect reduces the posterior probability of the null hypothesis to a low value. We show that this intuition is not true in general. The common situation in psychology and other areas is that the null hypothesis is a priori quite likely to be true. In such a situation, contrary to intuition, finding a significant effect leaves us with much posterior uncertainty about the null hypothesis being true. Obtaining a reasonable reduction in uncertainty is thus another reason to adopt the recent recommendation by \cite{benjamin2018redefine} to change type-I error rate to $\alpha = 0.005$. Furthermore, conducting high power studies is an obvious but neglected remedy. 
Otherwise, the results will be indecisive, even when replication studies are performed. 

Our key result is that the posterior probability for the null given a significant effect varies widely across settings involving different type-I and type-II error rates and different prior probabilities for the null. The intuition that frequentist p-values may provide a shortcut to this information is in general misleading.

\begin{tcolorbox}[colback=green!5!white,colframe=green!75!black,title=Conclusion]
  The most conservative way to proceed is:
  \begin{enumerate}
    \item Assume conservatively that the null hypothesis has a relatively high probability of being true.
    \item Set alpha at 0.005 \citep{benjamin2018redefine}.
    \item Run as high-powered a study as possible. This requires planning sample sizes in advance \citep{SampleSizeCBB2022}.
    \item Replicate your result once at least to make sure you can get similar ballpark estimates - \cite{gelmanhill07} call this the ``secret weapon''.
    \item Optionally: Understand the design properties of your experiment under hypothetical repeated sampling \citep{gelmancarlin}.
    \item Optionally: To gain a better understanding of your inferences, compute (and perhaps also report) the posterior probability of the null hypothesis by using the Shiny app that we developed here or by using the formulas shown in the present work, and/or conduct formal Bayes factor analyses \citep{schad2022workflow}.
  \end{enumerate}
\end{tcolorbox}

\section*{Acknowledgements}

Thanks to Valentin Amrhein and Sander Greenland for comments. 
Funded by the Deutsche Forschungsgemeinschaft (DFG, German Research Foundation) – Project number 317633480, SFB 1287.

\section*{Author contribution}

SV had the idea for the paper. SV and DJS performed analyses. SV and DJS generated the shiny app. SV and DJS wrote the paper.

\section*{Availability of simulations and computer code}

All the computer code used for the simulations reported in the present manuscript, and the code for generating all Figures, will be freely available online at https://osf.io/9g5bp/. Moreover, we make a shiny app available at https://danielschad.shinyapps.io/probnull/ that allows computing the posterior probability for the null given a significant effect for many different settings of type-I and -II error rates and for different prior probabilities for the null.

\bibliographystyle{apacite}
\bibliography{bibcleaned}

\end{document}